\documentclass[10pt,journal,a4paper,twoside,twocolumn]{IEEEtran}

\usepackage{amsthm,amsmath,amssymb}
\usepackage{psfrag}
\usepackage{pifont}
\usepackage{graphicx}
\usepackage[nocompress]{cite}
\graphicspath{{drawings/}{pics/}}
\usepackage{color}

\usepackage{algorithmic}
\usepackage[ruled]{algorithm}

\def\ve#1{{\mathchoice{\mbox{\boldmath$\displaystyle #1$}}%
              {\mbox{\boldmath$\textstyle #1$}}%
              {\mbox{\boldmath$\scriptstyle #1$}}%
              {\mbox{\boldmath$\scriptscriptstyle #1$}}}}
\DeclareSymbolFont{AMSb}{U}{msb}{m}{n}
\DeclareSymbolFontAlphabet{\mathbb}{AMSb}
\def\R{\mathbb{R}}

\def\sign{\mathrm{sign}}
\def\VA{\mathsf{VA}}
\newcommand {\dx} {~\mathrm{d}}
\def\GLRT{\mathsf{MSDD}}

\newcommand {\Ti} {T_{\mathsf{i}}}     
\newcommand{\T}{T}
\newcommand{\Nsym}{N}
\newcommand{\bDFDD}{{\mbox{\textsf{\scriptsize bDF-DD}}}}
\newcommand{\cDFDD}{{\mbox{\textsf{\scriptsize cDF-DD}}}}
\newcommand{\sbDFDD}{{\mbox{\textsf{\scriptsize sbDF-DD}}}}
\newcommand {\Rstop} {R_{\mathsf{stop}}} 
\def\dB{\mathrm{dB}}
\def\ns{\,\mathrm{ns}}
\def\GHz{\,\mathrm{GHz}}
\newcommand {\Eb} {E_{\mathrm{b}}}     
\def\No{N_0}

\def\EbNodB{10\log\left({\Eb}/{\No}\right)}
\def\BER{\mathrm{BER}}

\newcommand{\argmin}{\mathop{\mathrm{argmin}}}
\newcommand{\argmax}{\mathop{\mathrm{argmax}}}

\renewcommand{\algorithmicif}{\textbf{{if}}}

\renewcommand{\algorithmicwhile}{\textbf{{{while}}}}

\definecolor{commentgray}{rgb}{.3,.3,.3}

\definecolor{highlightgray}{rgb}{.9,.9,.9}

\newcommand{\IFONELINE}[2]{\STATE \algorithmicif ~ #1 {\{} #2 {\}}}
\newcommand{\WHILEONELINE}[2]{\STATE\algorithmicwhile ~ #1 {\{} #2 {\}}}

\DeclareFontFamily{OT1}{phv}{}
\DeclareFontShape{OT1}{phv}{m}{n}{ <-> [0.95] phvr8t }{}

\title{Decision-Feedback Differential Detection in\\Impulse-Radio Ultra-Wideband Systems}
\IEEEoverridecommandlockouts
\author{
\IEEEauthorblockN{Andreas~Schenk,~\IEEEmembership{Student~Member,~IEEE}, and Robert~F.H.~Fischer,~\IEEEmembership{Senior~Member,~IEEE}}
\IEEEauthorblockA{}
\thanks{This work was supported by the Deutsche Forschungsgemeinschaft (DFG) within the framework UKoLoS under grant FI 982/3-1.}
\thanks{Andreas~Schenk and Robert~F.H.~Fischer are with the Lehrstuhl f{\"u}r Informations{\"u}bertragung, Universit{\"a}t Erlangen--N{\"u}rnberg, Erlangen, Germany, email: \texttt{\{schenk,fischer\}@lnt.de}}
}
\begin{document}
\maketitle
\begin{abstract}
In this paper we present decision-feedback differential detection (DF-DD) schemes for auto\-cor\-re\-la\-tion-based detection in impulse-radio ultra-wideband (IR-UWB) systems, a signaling scheme regarded as a promising candidate in particular for low-complexity wireless sensor networks.
To this end, we first discuss ideal noncoherent sequence estimation and approximations thereof based on block-wise multiple-symbol differential detection (MSDD) and the Viterbi algorithm (VA) from the perspective of tree-search/trellis decoding.
Exploiting relations well-known from tree-search decoding, we are able to derive the novel decision-feedback differential detection (DF-DD) schemes.
A comprehensive comparison with respect to performance and complexity of the presented schemes in a typical IR-UWB scenario reveals---along with novel insights in techniques for complexity reduction of the sphere decoder applied for MSDD---that sorted DF-DD achieves close-to-optimum performance at very low, and in particular constant receiver complexity.
\end{abstract}
\section{Introduction}
\label{sec:intro}
\IEEEPARstart{U}{ltra-wideband} (UWB) transmission systems are widely regarded as a promising technique for short-range applications like wireless sensor networks (WSNs) \cite{UWB:Zhangetal:UWBWirelessSensorNetworks}, as the relatively large signaling bandwidth enables a reduced transmit power spectral density, coexistence to established narrow-band systems, and supports a large number of simultaneous users.
In particular, impulse-radio UWB (IR-UWB) is especially well suited for WSNs due to its robustness to severe multi-path fading even in indoor environments, the potential to provide accurate localization, and, last but not least, due to its low cost and complexity \cite{UWB:Win:IR}.
Moreover, commonly in WSNs information is transmitted in relatively short bursts and only low data rates have to be supported, such that intersymbol interference can easily be avoided.
We denote the burst length by $N$.

Avoiding costly channel estimation, low-complexity IR-UWB receivers rely on noncoherent detection  such as en\-er\-gy detection in the case of pulse-position-mod\-u\-lated IR-UWB, or autocorrelation detection in the case of pulse-amplitude-modulated IR-UWB \cite{UWB:Witrisaletal:NoncoherentUWBSystems}, cf., e.g., (differential) transmitted-reference \cite{UWB:Chao:TR}.
In particular, an autocorrelation receiver (ACR) enables conventional symbol-wise differential detection (DD) \cite{UWB:Chao:TR}.
The performance of ACR-based DD (in terms of the signal-to-noise ratio (SNR) to guarantee a desired bit error rate (BER)), however, suffers a large gap compared to idealistic detection assuming perfect channel estimation.
This gap can be bridged to a large extend, when jointly deciding for the best sequence of the $N$ symbols within the burst based on correlations of the receive signal ranging over the entire burst interval, i.e., employing an $N$-branch ACR \cite{UWB:Guo:MSDD}.
However, for large burst length, this ideal noncoherent sequence estimation imposes a very high complexity burden, as it a) requires delaying and correlating the receive signal over the entire burst interval, and b) exhibits a very high computational complexity to find the optimum sequence (exponential in $N$).
Hence, reduced-complexity detection schemes---still achieving close-to-optimum performance---are requested.

In this paper, we focus on detection schemes employing a---still extended, but reduced-complexity---$L$-branch ACR, where $L\ll N$.
We first review two well-known techniques, and show how both are connected to ideal noncoherent sequence estimation from the perspective of tree-search decoding \cite{TS:Gamal:TS}.
In particular, we consider block-wise multiple-symbol differential detection (MSDD) \cite{UWB:Guo:MSDD} employing the sphere decoder (SD) \cite{UWB:Lottici:MSDD} in combinations with techniques for complexity reduction, e.g., \cite{me:ICUWB09}, and detection based on the Viterbi algorithm (VA) \cite{UWB:Lottici:MSDD}.
The drawback of both methods, despite of their good performance, is that their computational complexity is relatively high and in the worst case increases exponentially with the blocksize or memory length, respectively.
The main contribution of this paper is to exploit the well-known relation of decision-feedback detection as an approximation of tree-search decoding \cite{TS:Gamal:TS,TS:Qureshi:RSSE,TS:DuelHeegard:DDSE}.
In doing so, we are able to transfer the concept of decision-feedback differential detection (DF-DD) \cite{MSDD:LeibPasupathy:DFDDAWGN,MSDD:Edbauer:DFDDAWGN,MSDD:Huber:DFDD} to ACR-based detection of IR-UWB, yielding a computational complexity only linear in $L$.
Similar approaches have successfully been applied, e.g., in the area of differential space-time modulation \cite{MSDD:PauliHuberLampe:DFSMSDD,MSDD:ZhuYiuSchober:NoncoherentRxDSTM}.
As known from multi-antenna systems, the performance of decision-feedback detection can be improved when decisions are taken in an optimized order \cite{MIMO:Foschini:BLAST}.
A comparison with respect to performance and complexity of the presented schemes allows us to conclude that such sorted variants of DF-DD for IR-UWB achieve close-to-optimum performance at very low, and in particular constant receiver complexity, thus realize  a very good performance-complexity tradeoff.

Noteworthy, besides ACR-based detection there are other promising non-autocorrelation-based approaches to IR-UWB detection, such as the related approaches based on a decision-directed ACR \cite{UWB:ZhaoLiuTian:DecisionDirectedACR} and on crosscorrelations with iteratively generated reference templates \cite{UWB:ZhouMaLottici:FastMultiSymbolBasedIterativeDetectors}, or approaches exploiting the sparsity of the UWB propagation channel via compressed sensing \cite{UWB:OkaLampe:CompressedSensing,UWB:ParedesArceWang:CompressedSensingCHEST} or RAKE reception employing a reduced number of fingers \cite{UWB:Lottici:CHEST}.

This paper is organized as follows: in Sec.~\ref{sec:system} the system model of IR-UWB and the ACR front-end are described.
The discussion of ACR-based detection schemes in Sec.~\ref{sec:detection} starts with ideal noncoherent sequence estimation, followed by two approximate methods based on MSDD and the VA, and is concluded with the presenation of the novel DF-DD schemes.
A summary of the complexity of these detectors allows us to conduct a comparison of the presented schemes in Sec.~\ref{sec:comparison}.
We conclude with final remarks in Section \ref{sec:last}.
\section{IR-UWB System Model}
\label{sec:system}
\subsection{Receive Signal Model}
\label{sec:rxsignal}
Throughout this paper, we consider transmission of binary pulse-amplitude-modulated IR-UWB in bursts of $\Nsym$ information symbols.
The receive signal is then given as
\begin{align}
        r(t) = \sum_{i=0}^{\Nsym} b_i p (t-i\T) + n(t)\;,
        \label{eq:rxsignal}
\end{align}
where $b_i\in\{\pm1\}$, $i=0,...,\Nsym$, are $\Nsym+1$ transmit symbols, which represent $\Nsym$ encoded information symbols $a_k\in\{\pm1\}$, $k=1,...,\Nsym$, and $\T$ is the symbol duration.
Differential encoding is assumed, such that $b_i = b_{i-1}a_i = b_0 \prod_{k=1}^{i}a_k$, with the reference symbol $b_0 = 1$, but equivalent encoding rules, e.g., multiple-symbol transmitted-reference (MSTR) \cite{UWB:ZhouMaRice:ISIT2010}, are also possible.
The overall receive pulse shape $p(t)$ results from the convolution of transmit pulse, receive filter, and channel impulse response;
its energy is normalized to one, thus, the energy per bit\footnote{%
Note that the energy for the first reference symbol is neglected, as typically relatively long bursts are considered.
} is given by $\Eb = 1$.
$n(t)$ is white Gaussian noise of two-sided power-spectral density $\No/2$, band-limited by the receive filter.
To preclude intersymbol interference, the symbol duration $\T$ is chosen sufficiently large, such that each pulse has decayed before the next pulse is received.

Note that the problem of timing acquisition and the usually applied frame structure used for time-hopping and code-division multiple access \cite{UWB:Win:IR,UWB:Win:TH} are not explicitly taken into account, as the latter 
can be regarded as additional linear block coding, or averaged out prior to further receive signal processing, cf., e.g., \cite{UWB:Lottici:MSDD,UWB:Witrisaletal:NoncoherentUWBSystems}.

\subsection{Autocorrelation-Based Detection}
\label{sec:ACR}
\begin{figure}[!t]
\centering
\includegraphics[width = 78mm]{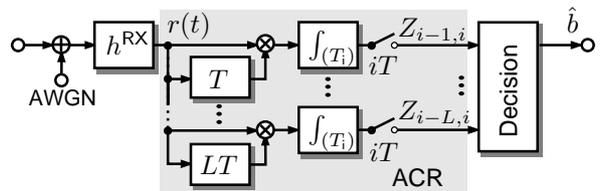}
\caption{Block diagram of an $L$-branch ACR.}
\label{fig:ACR}
\end{figure}

The core-ingredient of all investigated schemes for IR-UWB signal detection is the analog (or sufficiently sampled) front-end depicted in Fig.~\ref{fig:ACR}, the so-called $L$-branch ACR.
For the $i$-th symbol interval, it computes the correlation coefficients ($l = 1,...,L$)
\begin{align}
        Z_{i-l,i} &= \int_0^{\Ti} r(t+(i-l)T)\,r(t+iT)\dx{t}
\label{eq:ACRoutput}
\end{align}
of the receive signal in the $i$-th and the $L$ preceeding symbol intervals.
The integration interval $\Ti$ ($\leq\T$) of the ACR is a receiver parameter, which can be adapted to the channel characteristics at hand (cf., e.g., \cite{UWB:Witrisaletal:NoncoherentUWBSystems}).

Demanding the channel to remain constant over an interval of $L+1$-symbols, an $L$-branch ACR provides information on the relation of the current symbol to the preceeding $L$ symbols.
The phase transition from $b_{i-l}$ to $b_i$ is superposed by an ``information $\times$ noise'' and ``noise $\times$ noise'' term, i.e.,
\begin{align}
        Z_{i-l,i} &= b_{i-l}b_i \int_{0}^{\Ti} p^2(t)\dx t + \eta_{i-l,i}
        \label{eq:ACRoutput_detail}
\end{align}
where $\eta_{i-l,i}$ collects all terms corrupted by noise.

Difficulties in hardware implementation of the ACR may be regarded as a question of technology.
In particular the realization of accurate analog delay lines remains a demanding task, cf., e.g., \cite{UWB:Witrisaletal:NoncoherentUWBSystems,UWB:Franz:TR}, but advances in speed of A/D converters \cite{ADC:Walden:surveyanalysis,UWB:Franz:TR} will soon solve this problem.

\section{Signal Detection}
\label{sec:detection}
In the case of a single-branch ACR ($L=1$), using $a_i = b_i b_{i-1}$, from (\ref{eq:ACRoutput_detail}) it can be seen that symbol-wise DD is performed, such that the information symbols are directly obtained as $a_i^{\mathsf{DD}} = \sign(Z_{i-1,i})$.
In the case of an extended ACR ($L>1$), there are various methods how to finally decide the transmit symbols based on the ACR output; these are described in the following.
Beginning with ideal noncoherent sequence estimation (INSE), we discuss approximations thereof based on MSDD and the VA.
From this detailed, but unified treatment, we are not only able to straightforwardly derive novel DF-DD schemes for IR-UWB detection under \ref{sec:dfdd}, exploiting relations well-known from tree-search decoding \cite{TS:Gamal:TS,TS:DuelHeegard:DDSE,TS:Qureshi:RSSE}, but also to conduct a comprehensive comparison with respect to performance and complexity in Sec.~\ref{sec:comparison}.
This summarizes, along with novel insights in techniques for complexity reduction of the SD for MSDD, the main contribution of the paper.

\subsection{Ideal Noncoherent Sequence Estimation (INSE)}
\label{sec:optNSE}
First, recall that ideal noncoherent sequence estimation (INSE) would jointly decide for the best sequence of $\Nsym+1$ symbols, taking into account the receive signal in the entire burst interval $0\leq t\leq(\Nsym+1)\T$.
As the statistics of the receive pulse shape $p(t)$ are unknown, according to generalized-likelihood ratio testing (GLRT) an explicit optimization over this unknown parameter is included \cite{book:Kay:statisticalsignalprocessing2}.
Collecting the transmit symbols in a vector, this results in solving \cite{UWB:Lottici:MSDD,me:IZS2010}
\begin{align}
        \ve{b}^{\mathsf{INSE}} = \argmax_{\tilde{\ve{b}}\in\{\pm1\}^{\Nsym+1},\; \tilde{b}_{0} = 1}\,\sum_{i=1}^{\Nsym}
              \left(\tilde{b}_i\sum_{l=0}^{i-1}  \tilde{b}_l\, Z_{l,i}\right)
\label{eq:optNSE}
\end{align}
where the statistics $Z_{l,i}$, $i = 1,...,\Nsym$, $l=0,...,i-1$, are obtained from an $\Nsym$-branch ACR.
It is evident that, for large $\Nsym$, this is infeasible for two reasons:
a) correlations of the receive signal over time delays of $\Nsym\T$ have to be performed, which requires accurate delay lines over possibly hundreds of symbols, making hardware implementation impossible, and
b) the computational complexity of finding the best sequence is exponential in $N$, and thus intractable, since---at least at worst case---any algorithm must perform an exhaustive search over all $2^{\Nsym}$ possible sequences.

Hence, methods which are based on an ACR with only $L\ll\Nsym$ branches and reduced computational complexity---at best linear in $L$---are called for.
To motivate these techniques, note that INSE for binary signaling can be viewed as a tree search problem in a binary tree of depth $N$, with the optimum sequence given by the path from the root to the leaf with maximum path metric.

\subsection{Multiple-Symbol Differential Detection (MSDD)}
\label{sec:msdd}
One possible approximation of the INSE search problem is to split the binary tree of depth $N$ into smaller subtrees of depth $L$ 
and to solve each one independent.
This corresponds to splitting the burst of $\Nsym$ symbols into smaller blocks of $L$ symbols, and perform block-wise MSDD, based on the receive signal only in the corresponding interval.
To this end, the burst of $\Nsym+1$ receive symbols is decomposed into $\Nsym/L$ blocks of $L+1$ symbols $\ve{b}_\kappa = [b_{\kappa L}, b_{\kappa L +1},...,b_{\kappa L + L}]$, $\kappa = 0,1,...,\Nsym/L$, each representing $L$ information symbols, which, due to the differential encoding, overlap by one symbol\footnote{If necessary the final block length is reduced to $L_{\mathsf{f}} = \Nsym\mod L$.}.

The decision metric of block-wise MSDD is directly obtained from the INSE decision metric when restricting it to the corresponding block intervals.
However, to facilitate the application of tree-search decoding algorithms, a constant is subtracted similar to \cite{UWB:Lottici:MSDD}, yielding, with $\argmax(x) = \argmin(-x)$, exemplarily for the first block,
\begin{align}
        \ve{b}^\GLRT_0 = \argmin_{\tilde{\ve{b}}\in\{\pm1\}^{L+1},\; \tilde{b}_{0} = 1}\, \sum_{i=1}^{L}
  \left( \sum_{l=0}^{i-1} \left(|Z_{l,i}| - \tilde{b}_i\, \tilde{b}_l\,Z_{l,i} \right)\right)\;.
\raisetag{4mm}\label{eq:MSDD}
\end{align}
The required statistics to solve (\ref{eq:MSDD}), $Z_{l,i}$, can be obtained from an ACR with only $L$ $(\ll N)$ branches.

Due to the reformulation, the $i$-th increment of the decision metric,
\begin{align}
        \delta_i = \sum_{l=0}^{i-1} \left(|Z_{l,i}| - \tilde{b}_i\, \tilde{b}_l\,Z_{l,i}\right)
\label{eq:MSDD:branchM}
\end{align}
is always non-negative and solely depends on the $i$ preceeding symbols $\tilde{b}_l$, $l=0,1,...,i-1$.
This allows to check the decision metric componentwise, and thus fits into the framework of general tree-search decoding and in particular enables the application of the SD \cite{SD:Agrell:Closest,UWB:Lottici:MSDD}.
Note there are further options to approximately solve (\ref{eq:MSDD}) efficiently, e.g., based on relaxations of the search problem\footnote{\label{comment_xiaoli}These approaches are not considered in the latter comparison, as they require operations of significantly higher complexity compared to the presented schemes, such as solving a semi-definite program in \cite{UWB:ZhouMaRice:ISIT2010} or calculating dominant eigenvectors in \cite{UWB:ZhouMaRice:NearOptimalMSDD}.} \cite{UWB:ZhouMaRice:ISIT2010,UWB:ZhouMaRice:NearOptimalMSDD}.

Clearly, for $L=\Nsym$, INSE is obtained;
for the case of $L=1$, i.e., the decision of a single information symbol, block-wise MSDD reduces to traditional DD.
In \cite{MSDD:PauliLampe:TSMSDD} it has been observed that using blocks overlapping by more than one symbol, so-called subset MSDD, yields further gains in performance at the cost of complexity. Due to lack of space this is not considered in this paper.

We briefly review SD-based MSDD%
\footnote{cf.\ \cite{UWB:Lottici:MSDD} for details, but note that in contrast to \cite{UWB:Lottici:MSDD} the presented SD operates on the transmit symbols $b_i$ rather than on the data symbols $a_i$, yielding certain benefits as described below.}:
employing the Schnorr\--Euchner search strategy, at some node at depth $i-1$ the SD chooses the branch labeled by $\hat{b}_i$ with minimum branch metric.
As the SD operates on the transmit symbols, using (\ref{eq:optNSE}) and (\ref{eq:MSDD:branchM}), this is directly given as
\begin{align}
        \hat{b}_i &= \argmin_{\tilde{b}_i \in \pm1}\,\delta_i= \argmax_{\tilde{b}_i \in \pm1}\,\tilde{b}_i\sum_{l=0}^{i-1}  \tilde{b}_l\,Z_{l,i}
= \sign \sum_{l=0}^{i-1} Z_{l,i} \tilde{b}_l\;.\raisetag{4mm}
\label{eq:MSDD:SchnorrEuchner}
\end{align}
The tree is only extended along this branch, if the partial decision metric $\sum_{\iota=1}^{i}\delta_\iota$ is less than the search radius $R$.
At the beginning this search radius can be chosen arbitrarily large, but is updated whenever a new (preliminary) best block is found.
The SD algorithm for MSDD of IR-UWB is summarized in pseudo-code representation in Fig.~\ref{fig:SDpseudocode} (including techniques for complexity reduction as described below).
For brevity we defined $\delta_i = q_i - \tilde{b}_i p_i$,
and the symmetric matrix $\ve{Z}\in\R^{L+1\times L+1}$ with elements $Z_{l,i} = Z_{i,l}$, $i,l=0,...,L$.
As it does not influence the decision we may force the diagonal elements to be $Z_{i,i} = 0$.
Exemplarily, for $L=2$ we have
\begin{align}
        \ve{Z}
  &= \left[\begin{matrix} 0 & Z_{0,1} & Z_{0,2}\\ Z_{0,1}& 0 & Z_{1,2}\\ Z_{0,2} & Z_{1,2} & 0 \end{matrix} \right]\;.
\label{eq:MSDD:Zmat}
\end{align}
\begin{figure}[!t]
%
%
\hrule
\vspace*{1mm}
$\ve{b}^\GLRT =\mbox{\texttt{MSDD}}( \ve{Z},\,\Rstop )$
\hrule
{\small
\begin{algorithmic}[1]
\STATE $R := +\infty$; $\Delta_0 := 0$
\STATE $b_0 := 1$; $i := 1$
\STATE $p_i := \sum_{l=0}^{i-1} Z_{l,i}\,b_l$; $q_i := \sum_{l=0}^{i-1} |Z_{l,i}|$
\STATE $b_i := \sign(p_i)$; $n_i := 1$
\WHILE{$i>0$}
        \STATE $\Delta_i := \Delta_{i-1} + q_i - b_i\,p_i$
        \IF{$\Delta_i < R$}
                \IF{$i < L$}
                        \STATE $i := i+1$ 
                        \STATE $ p_i := \sum_{l=0}^{i-1} Z_{l,i}\,b_l$; $q_i := \sum_{l=0}^{i-1} |Z_{l,i}|$
                        \STATE $b_i := \sign(p_i)$; $n_i := 1$
                \ELSE
                        \STATE ${\ve{b}}^\GLRT := \ve{b}$; $R:=\Delta_i$
                        \IFONELINE{$R < \Rstop$}{\textbf{break and return} ${\ve{b}}^\GLRT$}
                        \STATE $i:=i-1$ 
                        \WHILEONELINE{$n_i > 1$}{$i := i-1$}
                        \STATE $b_i := -b_i$; $n_i := n_i +1$ 
                \ENDIF
        \ELSE
                \STATE $i := i-1$ 
                \WHILEONELINE{$n_i > 1$}{$i := i-1$}
                \STATE $b_i := -b_i$; $n_i := n_i +1$
        \ENDIF
\ENDWHILE
\end{algorithmic}
}
\caption{Pseudo-code representation of the SD algorithm for MSDD of IR-UWB.}
\vspace*{-5mm}
\label{fig:SDpseudocode}
\end{figure}

We consider three techniques to speed up the SD search process, two of which are presented here, the third is presented along with DF-DD in Sec.~\ref{sec:bdfdd}:
\subsubsection{SD stopping radius}
\label{sec:rstop}
The search process is terminated early if the metric of any preliminary sequence during the SD search process is less than
a precomputed stopping radius $\Rstop$, cf., Line 14.
In \cite{me:ICUWB09} it has been shown that choosing the stopping radius as
\begin{align}
        \Rstop = L \cdot \mathop{\min_{i,l,\,i\neq l}}_{} |Z_{l,i} |
\label{eq:MSDD:rstop}
\end{align}
preserves the optimality of the SD output.

\subsubsection{Initial SD search radius}
\label{sec:rinit}
Instead of choosing the initial SD search radius arbitrarily large (cf., Line 1), it may be chosen to any good estimate.
This can, e.g., be the decision metric of the DD sequence. In this case, if the SD does not find a better sequence, $\ve{b}^\GLRT = \ve{b}^{\mathsf{DD}}$.
If already the metric of the DD sequence meets the above stopping criterion, a SD call is not necessary at all.

\subsection{Viterbi Algorithm}
\label{sec:VA}
Another technique to approximate INSE, suggested in \cite{UWB:Lottici:MSDD}, employs the Viterbi algorithm (VA).
Note that, in the case of INSE, the memory length increases linearly from $1$ to $\Nsym$.
To enable the implementation of the VA, the memory length is truncated to a maximum of $L$.
Again taking the view that INSE is a search in a binary tree of depth $N$, this results in nodes, which may be assumed to be equivalent starting from a depth greater than $L$.
Merging these nodes, a trellis structure with a total of $2^L$ states is obtained.
This procedure is in the spirit of delayed decision-feedback sequence estimation \cite{TS:Qureshi:RSSE,TS:DuelHeegard:DDSE}.
The path metric for the VA is obtained from the INSE metric (\ref{eq:optNSE}) by restricting the memory length to $L$, i.e.,
\begin{align}
        \Lambda^\VA (\tilde{\ve{b}})  &= \sum_{i=1}^{\Nsym} \left( \tilde{b}_i \sum_{l=\max(0,i-L)}^{i-1} \tilde{b}_l Z_{l,i}\right)\;.
\label{eq:VA}
\end{align}
As usual for the VA, the final estimate is the sequence with maximum path metric.
The VA ensures a fixed complexity (exponential in $L$, but linear in $\Nsym$).
Again, an $L$-branch ACR is sufficient.
Depending on $L$, the VA will tradeoff between DD ($L=1$) and INSE ($L=\Nsym$); thus, for $L=\Nsym$, MSDD and the VA are equivalent.

\subsection{Decision-Feedback Differential Detection (DF-DD)}
\label{sec:dfdd}
The VA, as well as block-wise MSDD in the worst case, have a complexity in the order of $2^L$, i.e., exponential in the memory length or blocksize, respectively.
It is desireable to have a complexity linear in $L$.
This can be achieved using the principle of decision-feedback differential detection (DF-DD) \cite{MSDD:LeibPasupathy:DFDDAWGN,MSDD:Edbauer:DFDDAWGN,MSDD:Huber:DFDD}.
There are essentially two variants of DF-DD for IR-UWB, both operating on the output of an $L$-branch ACR: block-wise DF-DD being closely related to block-wise MSDD, and continuous DF-DD being related to the VA implementation.

\subsubsection{Block-Wise DF-DD (bDF-DD)}
\label{sec:bdfdd}
Block-wise DF-DD is directly obtained from SD-based block-wise MSDD \cite{me:ICUWB09}.
The Schnorr-Euchner search strategy in the SD for MSDD ensures that the first estimate in the SD search process equals DF-DD.
Thus, terminating the SD after the first point found, results in DF-DD with a linearly increasing feedback window length (from $1$ to $L$).
This is achieved, e.g., by calling the SD with $\Rstop = \infty$, or, equivalently, choosing ${b}_{0}^{\bDFDD} = 1$, and, similar to (\ref{eq:MSDD:SchnorrEuchner}),
\begin{align}
        {b}_{i}^{\bDFDD}
 = \sign\sum_{l=0}^{i-1} Z_{l,i} {b}_{l}^{\bDFDD}\;.
\label{eq:DFDD:block}
\end{align}

It is well known---especially from DF equalization in multi-antenna systems, also known as BLAST \cite{TS:Gamal:TS,MIMO:Foschini:BLAST}---that taking the decisions in an optimized order, i.e., employing some sorting, improves the performance.
Similarly, in the context of IR-UWB, interchanging the decision order within a block is enabled through the block-wise processing of bDF-DD (then labeled sorted block-wise DF-DD (sbDF-DD)).
Interchanging the decision order can easily be achieved by reordering the columns and rows of $\ve{Z}$ acc.\ to some sequence $\langle\hat{i}_0,\hat{i}_1,...,\hat{i}_{L}\rangle$, $\hat{i}_k\in\{0,...,L\}$, $\hat{i}_k\neq\hat{i}_l$ for $k\neq l$.

A reasonable sorting criterion can be derived from the DF-DD process itself.
For reliable decisions in each step the magnitude of the argument of the $\sign$-function in (\ref{eq:DFDD:block}) is desired to be as large as possible.
Hence, with ${\hat{i}_0} = 0$, $b_{0}^\sbDFDD = 1$, the first decided symbol should be the $\hat{i}_1$-th symbol, where $\hat{i}_1 = \argmax_{i=1,...,L} |Z_{0,i}b_{0}^\sbDFDD|$.
Taking the previous decision into account, 
the symbol which can be decided most reliable next can be found successively from
\begin{align}
        \hat{i}_k &= \argmax_{i\in\{1,...,L\}/\{\hat{i}_1,...,\hat{i}_{k-1}\}} \left|\sum_{l=0}^{k-1} Z_{\hat{i}_l,i} b_{\hat{i}_l}^\sbDFDD\right|
\label{eq:DFDD:sort}
\end{align}
and its value reads
\begin{align}
{b}_{\hat{i}_k}^\bDFDD &= \sign (\sum_{l=0}^{k-1} Z_{\hat{i}_l,i} b_{\hat{i}_l}^\sbDFDD)\;,
\label{eq:DFDD:sortdec}
\end{align}
where $k=1,...,L$.
Basically, this sorting criterion forces reliable decisions for the first decided symbols, which then strongly influence the upcoming decisions%
\footnote{
Different sorting criteria are also possible, e.g., the $l_1$-/$l_{\infty}$-norm (column/row norm are equivalent) of the matrix $\ve{Z}$, or acc.\ to the first row of $\ve{Z}$. However, we have found that all show some loss compared to successive sorting during the DF-DD process (up to $1\,\dB$ for the $l_{\infty}$-norm and the first-row criterion, and only marginal loss for the $l_1$-norm).
}.
It has to be noted that in contrast to BLAST, sorting is done per block based on the actual receive symbols and taking the previous decisions into account, rather than on the channel realization.

Noteworthy, the special case of sorted block-wise DF-DD and $L=2$ is equivalent to MSDD, i.e., $\ve{b}^\sbDFDD = \ve{b}^\GLRT$.
To proof this, assume $\ve{Z}$ is sorted, 
thus $|Z_{0,1}| \geq |Z_{0,2}|$ holds (cf.\ (\ref{eq:MSDD:Zmat})).
Since sbDF-DD chooses ${b}_0^\sbDFDD = 1$, ${b}_1^\sbDFDD = \sign(Z_{0,1})$, and
\begin{align}
        {b}_2^\sbDFDD = \sign(Z_{0,2} + \sign(Z_{0,1})Z_{1,2})\,,\nonumber
\intertext{the MSDD metric evaluates to}
        \Delta = |Z_{0,2}| + |Z_{1,2}| - \left|Z_{0,2} + \sign(Z_{0,1}) Z_{1,2}\right|\;.\nonumber
\end{align}
If $\sign(Z_{0,2}) = \sign(Z_{0,1}Z_{1,2})$, $\Delta = 0$.
Otherwise, either $\Delta = 2|Z_{1,2}|$ or $\Delta = 2|Z_{0,2}|$, depending on whether $|Z_{1,2}|<|Z_{1,2}|$ or vice versa, respectively.
In any case the minimum possible MSDD metric $\Delta \in \{0, 2\min |Z_{l,i}|\}$ results.

A similar sorting step may also be used as a preprocessing step in the case of MSDD using the SD, however as no previous decisions are available, (\ref{eq:DFDD:sort}) is upper bounded using the triangular inequality.
Note that sorting is only possible as the SD operates on the transmit symbols.
Since the SD starts with the DF-DD sequence, using a sorted SD input---and correspondingly reordering the output---delivers an improved first preliminary sequence and correspondingly updated search radius, thus speeds up the SD search process.
Note that sorting of the SD input---among different preprocessing steps---is a well known technique for SD complexity reduction \cite{TS:Gamal:TS,SD:Agrell:Closest,SD:ZhaoGiannakis:Lattices}.

\subsubsection{Continuous DF-DD (cDF-DD)}
\label{sec:cdfdd}
Another variant of DF-DD with a symbol-wise processing (sliding window) can be derived from the VA making use of the relation of the VA and DF-DD, which has been established in \cite{TS:DuelHeegard:DDSE,TS:Qureshi:RSSE}; in this view DF-DD corresponds to a reduced-state sequence estimation with only a single state.
For cDF-DD of IR-UWB, in contrast to (\ref{eq:DFDD:block}), now $L$ previous decisions are fed back to improve the decision of the current symbol, thus, fixing the memory length to $L$ as in the VA.
In detail, cDF-DD chooses
\begin{align}
        {b}_i^\cDFDD = \sign \sum_{l=\max(0,\,i-L)}^{i-1} Z_{l,i} {b}^\cDFDD_l\;.
\label{eq:DFDD:cont}
\end{align}
Due to the symbol-wise processing, sorting is not applicable for cDF-DD.
The transient behavior at the beginning of the stream leads to cDF-DD and bDF-DD being equivalent when $L = \Nsym$.

\section{Comparison}
\label{sec:comparison}
In this section, we compare the presented IR-UWB detection schemes in terms of performance and complexity.
We first define the complexity measure adopted in this paper and then assess the performance-complexity tradeoff via numerical results.

\subsection{Complexity}
Since all schemes (apart from INSE, which only serves as a reference) are based on the output of the same $L$-branch ACR, we focus on the computational complexity of the decision unit. 
Due to binary signaling, all multiplications (e.g., in (\ref{eq:MSDD})) are limited to sign-inversions, thus, assuming a suitable number format, such that sign-inversion and the $\sign(\cdot)$- and $|\hspace*{-1mm}\cdot\hspace*{-1mm}|$-operation require negligible complexity (e.g., two's complement), the main source of computational complexity of SD-based MSDD, the VA, or variants of DF-DD is the number of real-valued additions (adds).

Due to the triangular structure, block-wise DF-DD performs $(L-1)/2$ adds per symbol, while continuous DF-DD requires $
(L-1)$ adds per symbol%
\footnote{For simplicity of implementation we neglect the edge effects in cDF-DD and the VA.}%
, both having a complexity linear in $L$.
The VA performs $2L$ adds per state, thus, in total $2L\cdot2^L$ adds per processed information symbol, yielding a complexity exponential in $L$.
Concerning SD-based MSDD, the number of real-valued adds of the SD search process depends on the realization of $\ve{Z}$ and in particular on the SNR.
It ranges from $2L\cdot2^{L}$ adds per block in the worst case%
\footnote{In the worst case the SD searches the entire tree. However, due to the Schnorr-Euchner search strategy, the involved metric calculations are performed in an efficient way, cf.\ Fig.~\ref{fig:SDpseudocode}, such that only $\sum_{i=1}^{L-1}2^i(\tfrac{1}{2}2+\tfrac{1}{2}(2(i-1)+2)) + 2^L\cdot\tfrac{1}{2}(2(L-1)+2) = L2^{L+1}$ adds are required.}%
~(per symbol exponential in $L$), to a minimum of $L(L+1)-1$ adds per block in the best case (per symbol linear in $L$).
The latter occurs when the first path found during the SD search process fulfills the stopping criterion, i.e., the SD only computes the decision metric of one particular sequence step by step.
The same number of adds is required to find an initial search radius for the SD based on the MSDD decision metric of a particular sequence.
In the case of sorted DF-DD, sorting does not add to the overall complexity, as sorting is done successively based on a similar expression as required for taken the decisions (arguments of (\ref{eq:DFDD:sort}) and (\ref{eq:DFDD:sortdec}) are equal).
However, if sorting is applied as a preprocessing step of the SD for MSDD, calculating the optimized order increases the complexity by $(L-1)/2-1$ adds per block of $L$ symbols.

With the argumentation above, the complexity of DD, of finding the stopping criterion (\ref{eq:MSDD:rstop}), and of the final differential decoding step to obtain the information symbols from the estimated transmit symbols may be neglected.

\subsection{Numerical Results}
\label{sec:numresults}
For all numerical simulations, a typical IR-UWB scenario has been considered: the transmit pulse shape is chosen as a Gaussian monocycle with $2.25\GHz$ center frequency and a bandwidth of $3.3\GHz$ (measured at $10\,\dB$),
the propagation channel is modeled acc.\ to IEEE-CM\,2 \cite{UWB:Molisch:Channel} (constant over the burst interval and each realization normalized to unit energy), and
the receive filter is matched to the transmit pulse shape.
We assume no intersymbol interference ($\T$ chosen sufficiently large), and, for this setting, $\Ti = 30\ns$ is a good compromise for the integration time of the ACR.
All results have been averaged over a large number of bursts.

\begin{figure}[!t]
\centering
\includegraphics[width = 88mm]{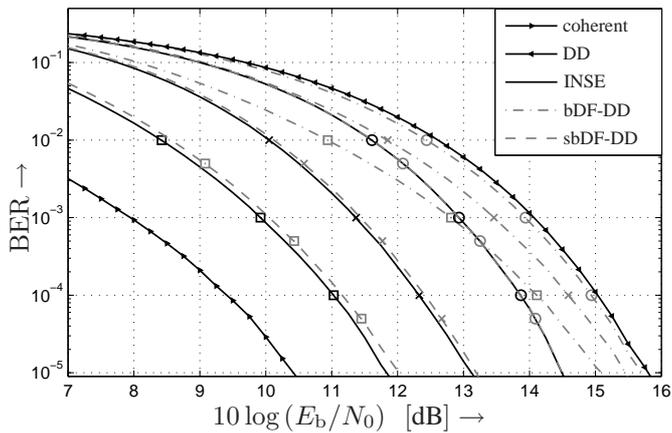}
\caption{$\BER$ performance vs.\ $\Eb/\No$ of DF-DD (sorted and non-sorted) in comparison to ideal noncoherent sequence estimation (INSE), DD, and ideal coherent detection of IR-UWB for short bursts ($\Nsym = L$) with $L=2$ (\textsf{o}), $L=5$ (\textsf{x}) and $L=15$ ({\tiny\textsf{$\square$}}).
IEEE-CM\,2, $\Ti = 30\,\ns$.}
\label{fig:BER_Nsmall}
\end{figure}

First, Fig.~\ref{fig:BER_Nsmall} depicts the BER for short bursts with $\Nsym = 2$, $5$, and $10$ symbols, where INSE can be realized by an $N$-branch ACR in combination with SD-based MSDD (all variants---sorted/non-sorted, with/without initial or stopping radius---have the same performance, and differ only in complexity).
INSE results in gains of about $4\,\dB$ over traditional DD for $N=15$.
Even for the relatively large feedback length of $L=\Nsym =15$, DF-DD (block-wise and continuous processing are equivalent for $L=\Nsym$) without sorting does not lead to significant gains vs.\ DD.
This is due to the linearly increasing feedback window length from $1$ to $L$, such that for the decision of the first decided symbols only few decisions are fed back.
The performance of DF-DD is tremendously improved, when the decision order is optimized as described under Sec.~\ref{sec:bdfdd}, yielding close-to-optimum performance for $L>2$, and, as shown under Sec.~\ref{sec:bdfdd}, exactly the same performance as MSDD, thus here also INSE, for the special case of $L=2$.

\begin{figure}[!t]
\centering
\includegraphics[width = 88mm]{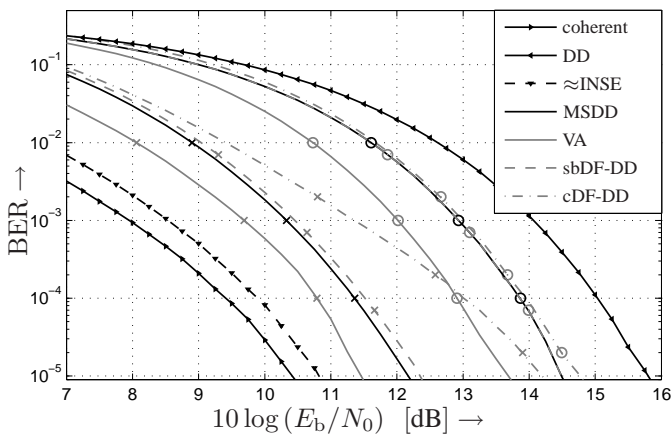}
\caption{$\BER$ performance vs.\ $\Eb/\No$ of DF-DD (block-wise with sorting and continuous) in comparison to block-wise MSDD, VA-based detection, DD, and ideal coherent detection of IR-UWB for bursts of $\Nsym = 100$ with $L=2$ (\textsf{o}), and $L=10$ (\textsf{x}). INSE is approximated by sorted DF-DD with $L=100$. IEEE-CM\,2, $\Ti = 30\,\ns$.}
\label{fig:BER_N100}
\end{figure}

In the case of a larger burst length ($\Nsym = 100$), as considered in Fig.~\ref{fig:BER_N100}, INSE becomes impracticable due to the high computational complexity and the required $N$-branch ACR.
For reference, due to the high complexity its performance is approximated by (sorted) DF-DD with $L\hspace*{.2mm}=\hspace*{.2mm}\Nsym\hspace*{.2mm} = 100$.
Naturally, the presented reduced-complexity detection schemes, employing only an $(L\ll\Nsym)$-branch ACR, i.e., block-wise MSDD, VA-based detection, and the DF-DD schemes, show increasing loss compared to INSE for decreasing $L$.
While block-wise MSDD (again all variants---sorted/non-sorted, with/without initial or stopping radius---show exactly the same performance) with a blocksize of $L$ is clearly outperformed by the continuous approach of VA-based detection with a fixed memory length of $L$, block-wise DF-DD with sorting and a linearly increasing memory length from $1$ to $L$ is superior to continuous DF-DD with a fixed memory length.
Thus, as known from other applications \cite{TS:Gamal:TS,MIMO:Foschini:BLAST}, in the case of DF-DD the sorting step, which is only applicable for block-wise processing, is crucial to achieve high performance with decision-feedback schemes.

%

\begin{figure}[!t]
\centering
\includegraphics[width = 88mm]{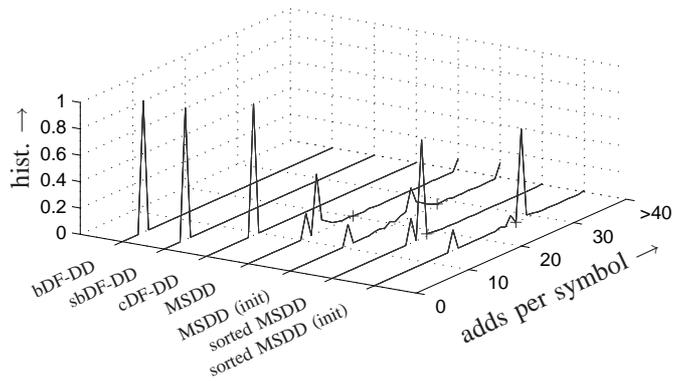}
\caption{Histogram of the complexity (in adds per symbol) of block-wise MSDD IR-UWB detection (with/without initial search radius, sorted/non-sorted), in comparison to block-wise (sorted/non-sorted) and continuous DF-DD at $\EbNodB = 10\,\dB$ for $L=10$.  Crosses: average complexity. IEEE-CM\,2, $\Ti = 30\,\ns$.}
\label{fig:hist}
\end{figure}

However, the VA---achieving best performance with an $L$-branch ACR---requires a significantly higher computational complexity compared to the other schemes, thus may be applied only in the case of very small $L$ (say, for $L\leq3$).
For $L = 10$,
Fig.~\ref{fig:hist} shows the complexity (measured as the number of adds per information symbol) of DF-DD and---due to the varying complexity---normalized histograms of the complexity of MSDD using the SD employing different combinations of the presented complexity reduction techniques (all use the packing-radius-based stopping radius, cf.\ (\ref{eq:MSDD:rstop}) and \cite{me:ICUWB09}) at an operating point of $\EbNodB =10\,\dB$, yielding a $\BER \approx 10^{-3}$.
The complexity of the VA is orders of magnitudes higher ($2L\cdot2^{L} = 20480$ adds per symbol for $L=10$) and is thus not included.
Straightforward application of the SD for MSDD, employing neither an initial search radius, nor sorting of the SD input, in many cases requires only relatively few additions (in the order of DF-DD), but there is a high variation, yielding a relatively large average, and very high worst-case complexity (cf.\ tails of the histograms with $>\hspace*{-1mm}40$ adds).
Surprisingly, incorporating an initial search radius based on DD mainly results in an increased complexity.
This is due to the fact that the increase in complexity of only calculating the MSDD decision metric of the DD sequence is not compensated by a sufficiently large search complexity reduction.
Although the sorting step prior to the SD adds to the overall complexity, as well, it is more than compensated afterwards, yielding reduced average complexity and significantly less variation.
Again, incorporating DD as an initial search radius mainly increases the complexity of sorted MSDD, such that we may conclude that MSDD employing a sorting step of the SD input and the packing-radius-based stopping criterion is the lowest-complexity variant among all SD-based variants for MSDD.
Similarly, sorted block-wise DF-DD is clearly preferable to other variants of DF-DD as it shows superior performance at half the complexity of continuous DF-DD and equal complexity as block-wise DF-DD without sorting.

%

\begin{figure}[!t]
\centering
\includegraphics[width = 88mm]{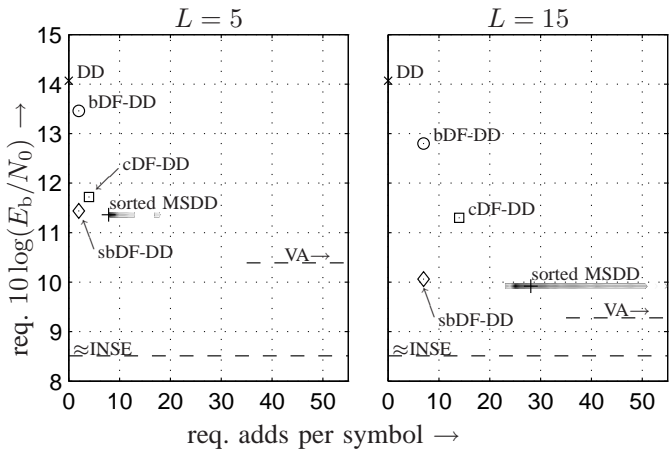}
\caption{Tradeoff performance vs.\ complexity at $\BER = 10^{-3}$ of IR-UWB detection using block-wise MSDD (sorted), block-wise (sorted and non-sorted) and continuous DF-DD, and the VA for different $L$ in comparison to DD and INSE. Only MSDD: histogram of complexity indicated as a colorbar (darker/lighter: higher/less occurrence, average complexity indicated by crosses). Complexity of VA is orders of magnitudes higher ($320$ and $15\cdot2^{16}$ adds per symbol for $L=5$ and $15$).  IEEE-CM\,2, $\Ti = 30\,\ns$.}
\label{fig:tradeoff}
\end{figure}

Finally, Fig.~\ref{fig:tradeoff} summarizes the tradeoff performance (in SNR to guarantee a desired $\BER$) vs.\ complexity (in adds per symbol) obtained with the presented schemes, i.e., DD and the more sophisticated schemes making use of an $L$-branch ACR (DF-DD, VA, and sorted block-wise MSDD), for $L=5$ and $L=15$, at an operating point of $\BER = 10^{-3}$ (thus compareable to $L=10$ at $\EbNodB = 10\,\dB$ in  Fig.~\ref{fig:hist}).
The fixed and average complexity is indicated by markers in the case of (DF-)DD and MSDD, respectively; for MSDD, a colormap also indicates the histogram of the complexity (darker/lighter: higher/less occurrence).
The complexity of the VA is orders of magnitudes higher than that of the other schemes ($320$ and $15\cdot2^{16}$ adds per symbol for $L=5$ and $15$, respectively); INSE (again approximated by sorted DF-DD with $L=N=100$) is not compareable in terms of complexity, as it requires an $N$-branch ACR;
thus, only the performance of both is indicated.

The detection schemes are lined up from lowest complexity and worst performance in the case of DD (no adds, $\mathsf{x}$), followed by block-wise ({$\mathsf{o}$}) and continuous ({\scriptsize $\square$}) DF-DD.
Further performance gains, at however higher, and in particular varying complexity, is achieved using sorted MSDD (average complexity, $+$).
The variation of MSDD complexity increases for increasing blocksize.
Employing an $L$-branch ACR, best performance at fixed, but very high complexity, is obtained using the VA.
The only exception to this strict line-up is block-wise DF-DD with an optimized decision order (sorted block-wise DF-DD, $\mathsf{\diamond}$), which achieves almost the performance of MSDD at significantly less complexity.

From this comparison we conclude that block-wise DF-DD in combination with sorting enables a very good performance-complexity tradeoff, a result which should be viewed in particular in comparison to other recently presented close-to-optimum block-based detectors, cf.\ \cite{UWB:ZhouMaRice:ISIT2010,UWB:ZhouMaRice:NearOptimalMSDD} and Footnote \ref{comment_xiaoli}.
Note that additionally scaling the stopping radius similar to \cite{me:ICUWB09} enables to smoothly switch between sorted MSDD and sorted DF-DD.

\section{Conclusions}
\label{sec:last}
In this paper we have presented autocorrelation-based decision-feedback differential detection (DF-DD) schemes for IR-UWB systems.
To this end, we reviewed multiple-symbol differential detection (MSDD) and detection based on the Viterbi algorithm in a unified way, frow which we derived the novel low-complexity DF-DD schemes, exploiting concepts well-known from tree-search decoding.
A comprehensive comparison with respect to performance and complexity of the presented schemes in a typical IR-UWB scenario reveals---along with new insights in techniques for complexity reduction of the sphere decoder applied for MSDD---that sorted DF-DD achieves close-to-optimum performance at very low, and in particular constant receiver complexity.

%
%
%
%


%
%
%
%

%
\end{document}